\documentclass[letterpaper]{JHEP3}

\pdfoutput=1

\usepackage{epsfig}
\usepackage{graphicx}
\usepackage{amssymb,amsmath}
\usepackage{epstopdf}

\def\be{\begin{eqnarray}}
\def\ee{\end{eqnarray}}

\title{Response of Holographic QCD to Electric and Magnetic Fields}

\author{Oren Bergman\thanks{On sabbatical leave from the Department of Physics,
Technion, Haifa 32000, Israel.}\\
School of Natural Sciences\\
Institute for Advanced Study\\
Princeton, NJ 08540, USA\\
\email{bergman@sns.ias.edu,bergman@physics.technion.ac.il}}

\author{Gilad Lifschytz\\
Department of Mathematics and Physics and CCMSC \\
University of Haifa at Oranim\\
Tivon 36006, Israel \\
\email{giladl@research.haifa.ac.il}}

\author{Matthew Lippert\\
Department of Physics\\
Technion, Haifa 32000, Israel\\
{\rm and}\\
Department of Mathematics and Physics \\
University of Haifa at Oranim\\
Tivon 36006, Israel\\
\email{matthewslippert@gmail.com}}

\date{}

\abstract{We study the response of the Sakai-Sugimoto 
holographic model of large $N_c$
QCD at nonzero temperature to external electric and magnetic fields.
In the electric case we find a first-order insulator-conductor transition in
both the confining and deconfining phases of the model.
In the deconfining phase the conductor is described by the parallel
8-brane-anti-8-brane embedding with a current of quarks and anti-quarks.
We compute the conductivity and show that it agrees precisely
with a computation using the Kubo formula.
In the confining phase we propose a new kind of 8-brane embedding,
corresponding to a baryonic conductor.
In the magnetic field case we show that the critical temperature for chiral-symmetry
restoration in the deconfined phase increases with the field and approaches a 
finite value in the limit of an infinite magnetic field. 
We also illustrate the nonlinear behavior of  the electric and magnetic 
susceptibilities in the different phases.}

\begin{document}

\section{Introduction and Summary}

Holographic descriptions of large $N_c$ QCD-like theories have 
received a great deal of attention recently (for a recent review see \cite{eekt}).
The models involve the physics of $N_f$ flavor-brane 
probes in near-horizon geometries of $N_c$ color-branes.
The Sakai-Sugimoto D4-D8-$\overline{\mbox{D8}}$
model \cite{Sakai_Sugimoto} in particular is very attractive in that
it has a simple geometric description of chiral symmetry breaking,
and appears to be rich enough to incorporate, at least qualitatively, all the low energy features
of QCD.
The properties of the mesons \cite{Sakai_Sugimoto,mesons} and  
baryons \cite{baryons}, the resolution of the $U(1)_A$ problem \cite{bl},
and the phase diagram at nonzero temperature \cite{asy,Parnachev:2006dn}, 
nonzero baryon chemical 
potential \cite{bll,baryon_density}, and nonzero isospin chemical potential \cite{isospin},
all exhibit many similarities with QCD. While the original model had
only massless quarks, it can be generalized to nonzero quark mass \cite{bss}.

The Sakai-Sugimoto model does not include a true electromagnetic gauge field,
but we can mimic the effect of one using the Abelian part of the flavor symmetry.
In this paper we will analyze the Sakai-Sugimoto model
at finite temperature with an Abelian background gauge field in the diagonal $U(1)_V$
part of the D8-$\overline{\mbox{D8}}$ gauge group.
This gauge field is holographically dual to the baryon number current
of the four-dimensional gauge theory.
Except for the one-flavor case, 
this is not the same as an electromagnetic field, since
``up" and ``down" type quarks have the same charge.
Nevertheless, we expect the physics to be qualitatively similar
to that of a background electromagnetic field.
Background flavor gauge fields have been studied previously in the 
${\cal N}=2$ supersymmetric theory corresponding to 7-brane
probes in the 3-brane background \cite{Karch,Johnson,Erdmenger}.

We would like to determine the effect of the electric and magnetic fields
on the phase diagram of the theory and compute the response 
coefficients (susceptibilities or conductivities) of the different phases.
We will do this by analyzing the 8-brane embeddings in the presence 
of the appropriate background world-volume gauge field.
The value of the 8-brane action is identified in the usual way with the 
appropriate thermodynamic potential.

In the case of an electric field we find an insulator-conductor type transition,
which in the deconfined phase generalizes the chrial symmetry breaking-restoration
transition at zero field. The critical temperature decreases with the field.
Since we use the DBI action, there is a temperature-dependent maximal value of the electric field.  
However, the transition to the conductor
occurs at a smaller value of the field.
We compute the conductivity in the conducting phase and show that it agrees
precisely, in the zero field limit, with what is expected from the Kubo formula.
This agreement extends to the nonzero density case as well.
Somewhat surprisingly, 
we also find an insulator-conductor transition in the confined phase,
where there is no chiral-symmetric phase.
We propose that the conducting phase corresponds to an embedding in
which the 8-brane and anti-8-brane are geodesically parallel and connect
at a cusp at the ``tip of the cigar". 
We compute the conductivity of this embedding and determine the phase diagram.
The current in this case is carried by baryons and anti-baryons.

With a magnetic field we observe that the critical temperature
for chiral symmetry restoration increases with the field,
in agreement with expectations for QCD.
We find that the critical temperature approaches a finite value in the limit
of infinite field, which differs from the behavior in the ${\cal N}=2$ theory
found in \cite{Johnson,Erdmenger}.

The paper is organized as follows:
In section 2 we review briefly the Sakai-Sugimoto model and set up our conventions;
sections 3 and 4 deal with, respectively, the physics of background electric and magnetic fields.

\section{Review of the Sakai-Sugimoto model}

The model consists of $N_c$ D4-branes in Type IIA string theory
wrapping a circle with anti-periodic boundary conditions for fermions, 
$N_f$ D8-branes at a point on the circle, and $N_f$
anti-D8-branes at another point on the circle. 
At energies well below the Kaluza-Klein scale the 
spectrum on the D4-branes is precisely that of massless four-dimensional
``QCD", with $N_c$ colors of gluons and $N_f$ flavors of quarks.
The holographic dual description comes about by taking the large $N_c$ limit,
in which the D4-branes are replaced by their near-horizon supergravity
background. It is most convenient to work in units in which the curvature radius
is fixed to unity, $R = \left(\pi g_s N_c\right)^{1/3} \sqrt{\alpha'} = 1$.
Taking $x_4$ as the circle direction, $x_4\sim x_4 + 2\pi R_4$,
the background is given by
\be
\label{SS_background}
ds^2 &=& u^{3\over 2} \left(-dx_0^2 + d{\bf x}^2
+ f(u) dx_4^2\right)
+ u^{-{3\over 2}}\left({du^2\over f(u)}
+ u^2 d\Omega_4^2\right) \nonumber\\
e^{\Phi} &=& g_s u^{3/4} \; , \;
F_4 = 3\pi (\alpha')^{3/2}  N_c \, d\Omega_4 \,,
\ee
where $f(u) = 1 - (u^3_{KK}/u^3)$, and $u_{KK} = 4/(9R_4^2)$.
The $(u,x_4)$ subspace is topologically a cigar (or disk),
with a tip at $u=u_{KK}$. In the dual gauge theory this ultimately implies
that the gluons are confined.
The gravitational description is valid as long as 
$\lambda \gg R_4$, where $\lambda$ is the five-dimensional 't Hooft
coupling, which in our units is given by
\be
\lambda = 4\pi g_s N_c \sqrt{\alpha'} = {4\over\alpha'} \,.
\ee

\begin{figure}[htbp]
\begin{center}
\begin{tabular}{ccc}
\epsfig{file=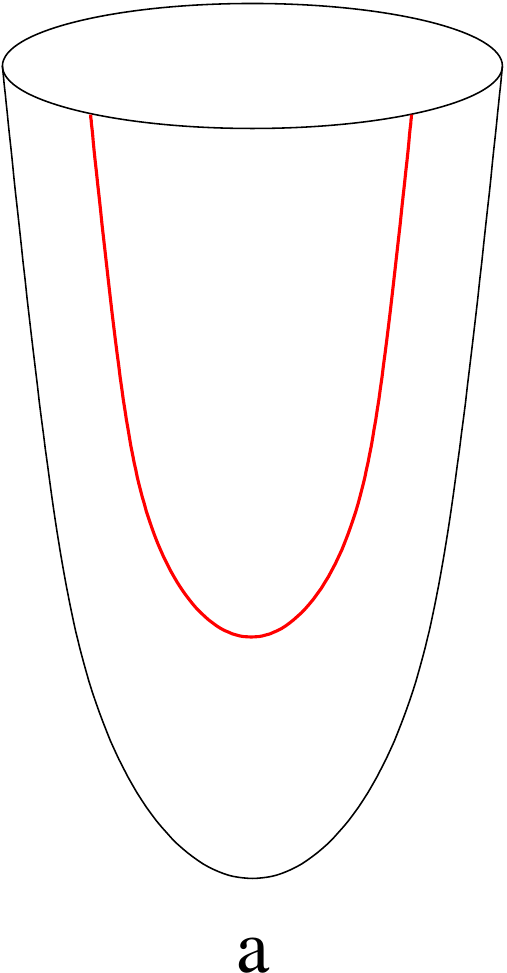,scale=0.4} \;\;\;\;\; &
\epsfig{file=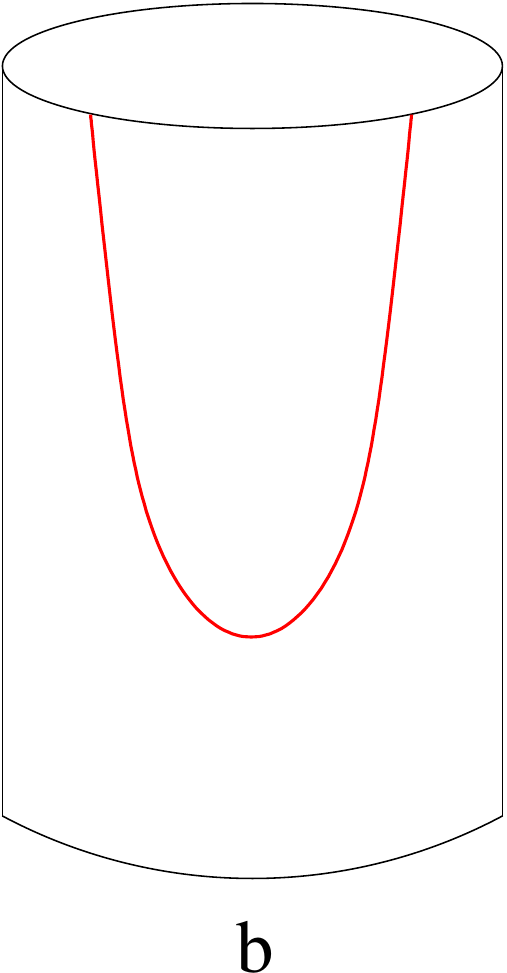,scale=0.4} \;\;\;\;\; &
\epsfig{file=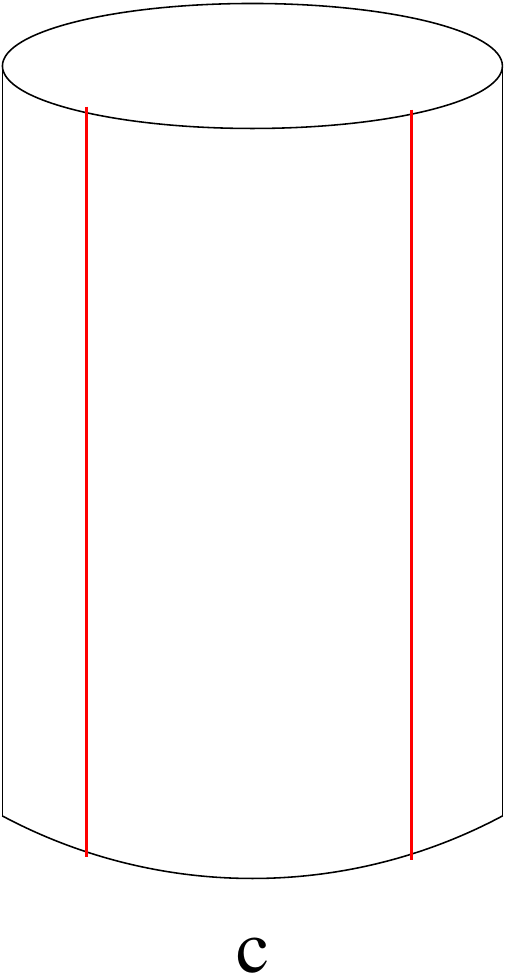,scale=0.4}
\end{tabular}
\caption{{\bf 8-brane embeddings and phases of the Sakai-Sugimoto model:
(a) confined, broken chiral symmetry (b) deconfined, broken chiral symmetry
(c) deconfined, restored chiral symmetry}}
\label{embeddings}
\end{center}
\end{figure}

The 8-branes and anti-8-branes are treated as probes in this background, and their embedding
determines the flavor physics in the dual gauge theory.  
Due to the topology of the background the 8-branes
and anti-8-branes must connect into a smooth U-shaped configuration at 
some radial position $u_0\geq u_{KK}$ (fig.~\ref{embeddings}a). 
This reflects the spontaneous breaking
of the $U(N_f)_R\times U(N_f)_L$ chiral symmetry to the diagonal $U(N_f)_V$.
The form of the embedding can be determined from
the DBI action of the 8-branes in this 
background,
\be
S = {\cal N} \int d^4x \int du \, u^4 \left[f(u)(x_4'(u))^2 + {1\over u^3f(u)}\right]^{1/2} \,.
\ee
The normalization constant is given by
\be
{\cal N} = 2N_f V_4 \Omega_4 T_8 = {4 N_f V_4\over 3(2\pi)^6 (\alpha')^{9/2} g_s} \,,
\ee
where $V_4$ is the volume of 4d spacetime,
$\Omega_4$ is the volume of a unit 4-sphere, and $T_8$ is the 8-brane tension.
The factor of 2 corresponds to the two halves of the embedding 
(8-branes and anti-8-branes) along $u$.
The equation of motion for the embedding $x_4(u)$ gives
\be
x_4'(u) = {1\over u^{3/2} f(u)} \left[{u^8f(u)\over u_0^8 f(u_0)} - 1\right]^{-1/2} \,.
\ee
Therefore at large $u$
\be
\label{large_u_embedding}
x_4(u) \approx {L\over 2} - {2\over 9} {c\over u^{9/2}} \,,
\ee
where $L$ is the asymptotic 8-brane-anti-8-brane separation,
\be
\label{L}
L = 2 \int_{u_0}^\infty du\, x_4'(u) \,,
\ee
and $c$ is the constant of the motion associated with $x_4(u)$.
This constant corresponds to the curvature of the 8-brane, and
is related to $u_0$ by
\be
\label{integration_constant}
c = u_0^4 \sqrt{f(u_0)} \,.
\ee

At nonzero temperature there are two possible backgrounds.
For $T<1/(2\pi R_4)$ the dominant background is the Euclidean continuation
of (\ref{SS_background}) with $x_0^E \sim x_0^E + 1/T$.
In this background the 8-brane embedding is the same as above.
The glue sector is therefore confined, and chiral symmetry is broken.
For $T>1/(2\pi R_4)$ the dominant background is given 
by (\ref{SS_background}), with the roles of $x_4$ and $x_0^E$ exchanged
and with $f(u)=1-(u_T^3/u^3)$, where $u_T=(4\pi T/3)^2$.
Here the $(u,x_4)$ subspace is topologically a cylinder,
with a horizon at $u=u_T$, which
in the dual gauge theory implies deconfinement.
In this background there are two possible 8-brane embeddings:
a U-shaped embedding (fig.~\ref{embeddings}b), that satisfies
\be
\label{U_deconfined}
x_4'(u) = {1\over u^{3/2} \sqrt{f(u)}} \left[{u^8f(u)\over u_0^8 f(u_0)} - 1\right]^{-1/2} \,,
\ee
and a parallel 8-brane-anti-8-brane embedding with $x_4'(u)=0$
(fig.~\ref{embeddings}c).
For $T<0.154/L$ the U embedding dominates and therefore
chiral symmetry is broken,
but when $T>0.154/L$ the parallel embedding
dominates and chiral symmetry is restored. The intermediate phase of 
deconfinement and chiral symmetry breaking appears only when
this critical temperature is higher than the deconfinement temperature $1/(2\pi R_4)$,
namely when $L<0.97 R_4$. 
Both the confinement/deconfinement and chiral symmetry breaking/restoration
transitions are first order.

\begin{figure}[htbp]
\begin{center}
\epsfig{file=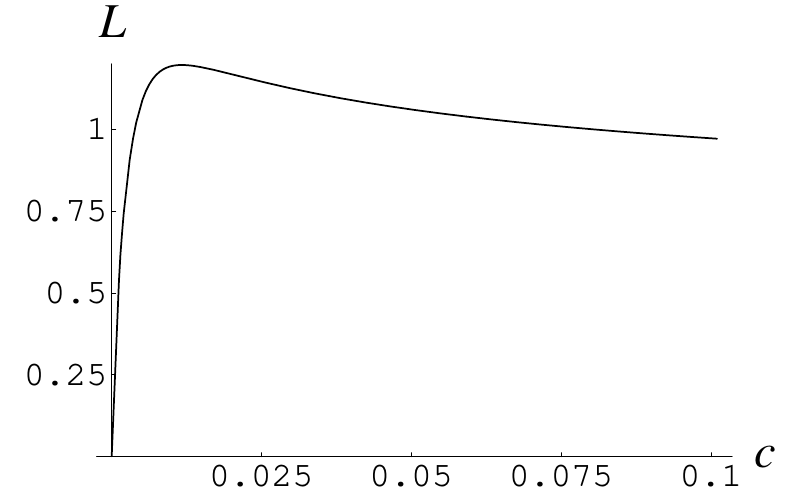, scale=0.7}
\caption{{\bf L vs. c for the U embedding in the deconfined phase ($T=0.14$)}}
\label{L_c_deconfined}
\end{center}
\end{figure}

The 8-brane curvature $c$ is an order
parameter for the chiral symmetry transition in the deconfined phase: 
it vanishes in the chiral-symmetric
parallel embedding, and is given by (\ref{integration_constant}) in the
chiral symmetry breaking U embedding.\footnote{This is not the usual chiral
symmetry order parameter. The 8-brane curvature $c$ is related to the expectation
value of a quark quadra-linear operator \cite{Antonyan:2006vw}.
The usual chiral symmetry order parameter is the quark bi-linear
$\langle\bar{q}q\rangle$, which is given by the normalizable mode
of the flavor bi-fundamental scalar field \cite{bss}.}
By studying the dependence of $L$ on $c$ one finds a turn-around
behavior typical of a first order phase transition.
This is easily seen from the asymptotic behavior of $L$ at small and large $c$
(see fig.~\ref{L_c_deconfined} for a numerical plot of $L$ vs. $c$ at a fixed temperature, and also \cite{Parnachev:2006dn}),
\be
\label{L_asymptotics}
\begin{array}{llll}
c \rightarrow 0 & (u_0\rightarrow u_T) & : & L(c,T)\sim c/T^9 \\
c\rightarrow \infty & (u_0\rightarrow \infty) & : & L(c,T)\sim c^{-1/8} \,.
\end{array} 
\ee
This implies that there is a maximal value of the asymptotic separation 
$L_{max}$, which depends on $T$.
For $L<L_{max}$ there are actually
two U embedding solutions, and for $L>L_{max}$ there are none.
Alternatively, since $L$ is a monotonically decreasing function of the temperature,
there is a maximal temperature $T_{max}$ at any fixed $L$.
For $T<T_{max}$ there are three solutions in all: the parallel embedding,
and the two U embeddings. Evidently, one of the U embeddings must be
an unstable solution. 
At $T=T_{max}$ the unstable U embedding merges with the stable one,
and at higher temperatures the two disappear, leaving only the parallel embedding.
The transition from the stable U embedding to the parallel embedding
occurs at a temperature lower than $T_{max}$.


\section{Electric Field}

In this section we will study the response of the model to an external electric field $E$,
by turning on an appropriate background value for the Abelian gauge field
component of the 
unbroken $U(N_f)_V$ gauge field in the 8-brane worldvolume.
We normalize this field as follows
\be
\hat{A} = {1\over N_f} \mbox{Tr}\,{\cal A}\,,
\ee
where ${\cal A}$ is the $U(N_f)_V$ gauge field.\footnote{The canonical
normalization is $\hat{A} = \sqrt{2/N_f}\, \mbox{Tr}\,{\cal A}$, but then
the quarks carry $1/\sqrt{2N_f}$ units of charge.
In our normalization the kinetic term has an extra factor of $2N_f$,
but the quarks carry unit charge. This normalization is more suitable to mimic
an electromagnetic field.}
Anticipating a current in the direction of the background field, we make the ansatz
(in Euclidean space)
\be
\label{electric_ansatz}
\hat{A}_1(x_0,u) = -iEx_0^E + F(u) \,,
\ee
where the $u$ dependence encodes the current in the usual holographic fashion \cite{Karch}.

\subsection{Deconfined phase}

We begin in the deconfining background, which dominates when $T>1/(2\pi R_4)$. 
The DBI action for the 8-branes
with the gauge field in (\ref{electric_ansatz}) is given by
\be
S = {\cal{N}} \int du \, u^4 \sqrt{\left(f(u) (x_4')^2 + \frac{1}{u^3} \right) 
\left(1- \frac{e^2}{f(u) u^3} \right)+ \frac{f(u)(a_1')^2}{u^3}} \,,
\ee
where we have defined the dimensionless quantities
$a_1 \equiv 2\pi\alpha' \hat{A}_1$ and $e \equiv 2\pi\alpha' E$.
The asymptotic behavior of the gauge field is given by
\be
a_1(x_0,u) \sim -iex_0^E - {2\over 3} {j\over u^{3/2}} \,,
\ee
where $j$ is the conserved charge associated with $a_1$,
namely the baryon number current. 
The physical dimensionful current is given in our units by
$J = (2\pi\alpha' {\cal N}/V_4) j$.
In terms of the current the action becomes
\be
\label{deconfined_action_e}
S = {\cal{N}} \int du \, u^4 \sqrt{\left(f(u) {x_4'}^2 + \frac{1}{u^3} \right) 
\left(f(u)- \frac{e^2}{u^3} \right)\left(f(u)-{j^2\over u^5}\right)^{-1}} \,.
\ee
This form of the action displays a generalization of the usual limiting
electric field of the flat space DBI action. 
For a vanishing current, the action is complex, and the embedding is unphysical,
if the second factor becomes negative anywhere in the integration region.
However this can be fixed by turning on a current such that the third
factor changes sign at the same point. This gives an equation
relating the current and the electric field \cite{Karch}.

As in the zero field case there are two types of embedding.
Consider first a U embedding with a vanishing current, $j=0$.
The solution satisfies
\be
\label{U_deconfined_e}
x_4'(u) = {1\over u^{3/2} \sqrt{f(u)}} 
\left[{u^8\left(f(u)-{e^2\over u^3}\right)
\over u_0^8 \left(f(u_0) - {e^2\over u_0^3}\right)} - 1\right]^{-1/2} \,,
\ee
and the large $u$ behavior is given by (\ref{large_u_embedding}), with
\be
\label{integration_constant_e}
c = u_0^4\sqrt{f(u_0) - {e^2\over u_0^3}} \,.
\ee
The action of this solution is given by
\be
\label{deconfined_U_solution}
S^{U} = {\cal{N}} \int_{u_0}^\infty  du  \, u^{5/2}
\sqrt{1-{e^2\over u^3 f(u)}}
\left[1 - {u_0^8\left(f(u_0)-{e^2\over u_0^3}\right)
\over u^8 \left(f(u) - {e^2\over u^3}\right)}\right]^{-1/2} \,.
\ee
The U embedding with $j=0$ is always physical:
since $c$ is real, the solution satisfies $e^2\leq u_0^3f(u_0)$,
and the action is real.
A current may be turned on, as long as $j^2<u_0^5 f(u_0)$,
but this increases the action, so the dominant U embedding has $j=0$.
In the gauge theory this embedding therefore corresponds to a chiral-symmetry
breaking, insulating phase. 

In fact the U embedding satisfies an even tighter bound on the electric field
than above. 
As in the zero field case, at fixed values of $e$ and $T$, there is a maximal
value of $L$ as a function of $c$ for the U embedding. 
Since $L$ is a monotonically decreasing function of $e$,
this implies a maximal value of $e$ as a function of $c$ at fixed values of $L$ and $T$.
The maximal value $e_{max}$ is attained at some $c>0$,
which, using (\ref{integration_constant_e}), implies that $e^2_{max}<u_0^3 f(u_0)$.
For $e>e_{max}$ there are no U embedding solutions, and we therefore
expect a phase transition to occur, at fixed $T$ and $L$, at some
value of $e$ smaller than $e_{max}$.


In the parallel embedding $x_4'(u)=0$, and the action is given by
\be
S^{||} = {\cal{N}} \int_{u_{T}}^\infty  du  \, u^{5/2} 
\sqrt{\frac{f(u) - {e^2\over u^3}}{f(u) - {j^2\over u^5}}} \,.
\ee
The numerator is negative for $u^3 < e^2 + u_T^3$, which is always in the range of integration.
The only way to ensure a real action in this case is for the denominator to become negative
at the same $u$. This requires a nonzero current given by
\be
j= e(e^2 + u_T^3)^{1/3} \,.
\ee
The parallel embedding therefore describes a chiral-symmetric conducting phase 
in the gauge theory, and the conductivity is given by
\be
\label{vacuum_conductivity}
\sigma = {J\over E}  = {(2\pi\alpha')^2{\cal N}\over V_4}\, (e^2 + u_T^3)^{1/3}  =
{N_f N_c\lambda T^2\over 27\pi} \left(1 + \tilde{e}^2\right)^{1/3} \,,
\ee
where we have defined a new dimensionless variable $\tilde{e}$ by
\be
\tilde{e} \equiv {e\over u_T^{3/2}} 
= {27\over 8\pi^2} {E\over \lambda T^3} \,.
\ee

To determine which phase dominates as a function of the temperature and
electric field we should in principle compare the electric free energies of the 
two solutions, which are in turn defined by the Euclidean 8-brane action of the solutions
\be
\label{electric_free_energy}
{\cal F}_e(L,e,T) = TS[x_4(u),a_1(u),T]_{EOM} \,.
\ee
However this is not quite right for the parallel embedding.
First of all, the conducting phase is not in equilibrium.
There is a steady state current with a finite conductivity,
meaning that energy must be constantly added to the system.  
This energy is dissipated into the gluon ``bath", which, in general, raises the temperature.  
At large $N_c$, however, this effect is negligible. While the dissipated energy
is ${\cal O}(N_c)$, there are ${\cal O}(N_c^2)$ gluons among which to distribute
this energy, so the temperature rise is only ${\cal O}(N_c^{-1})$.
But even ignoring the dissipation, one still needs to subtract the kinetic
energy of the current carriers, which should not be taken as part 
of the budget at the phase transition.

\begin{figure}[htbp]
\begin{center}
\epsfig{file=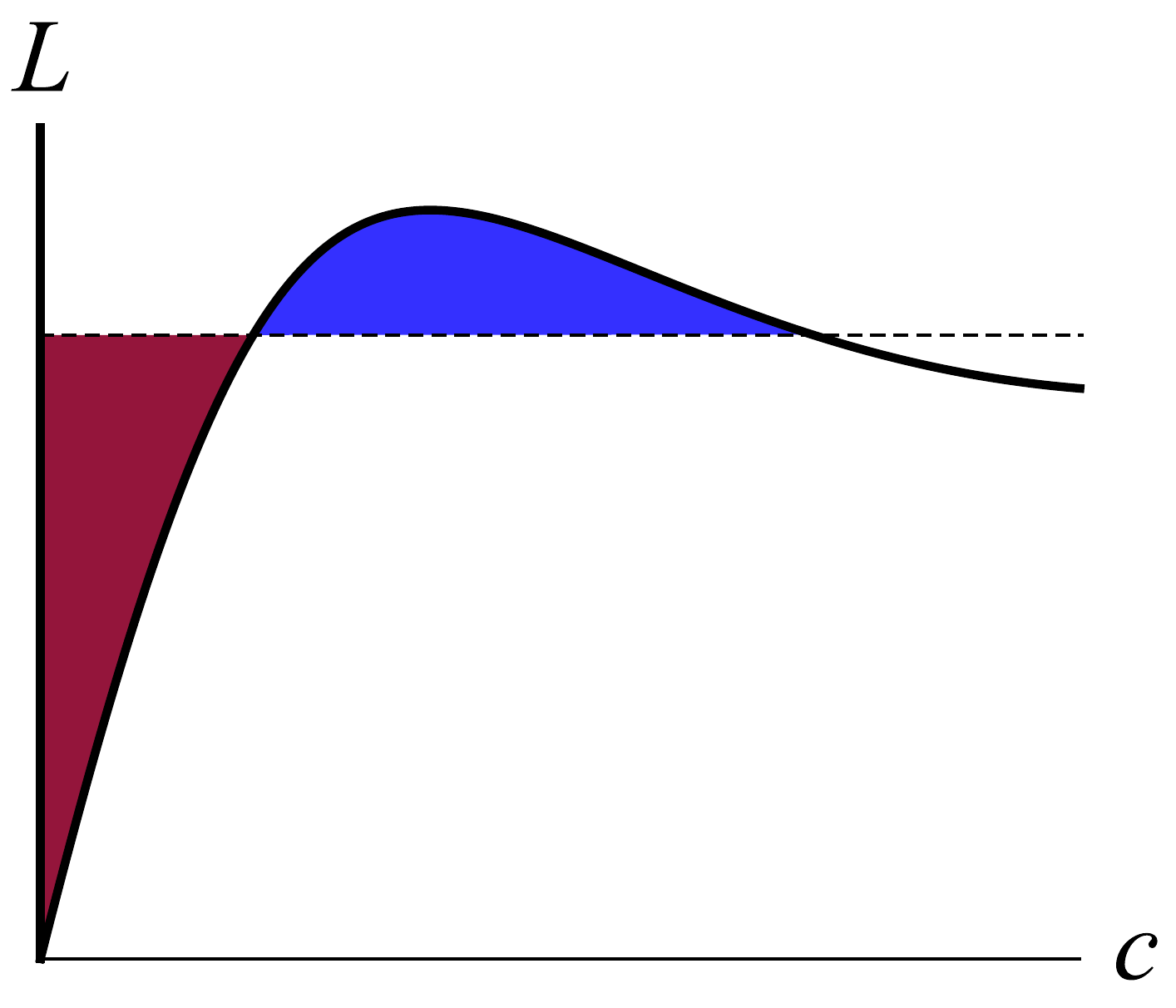, scale=0.3}
\caption{{\bf Illustration of the Maxwell construction}}
\label{Maxwell}
\end{center}
\end{figure}

\begin{figure}[htbp]
\begin{center}
\epsfig{file=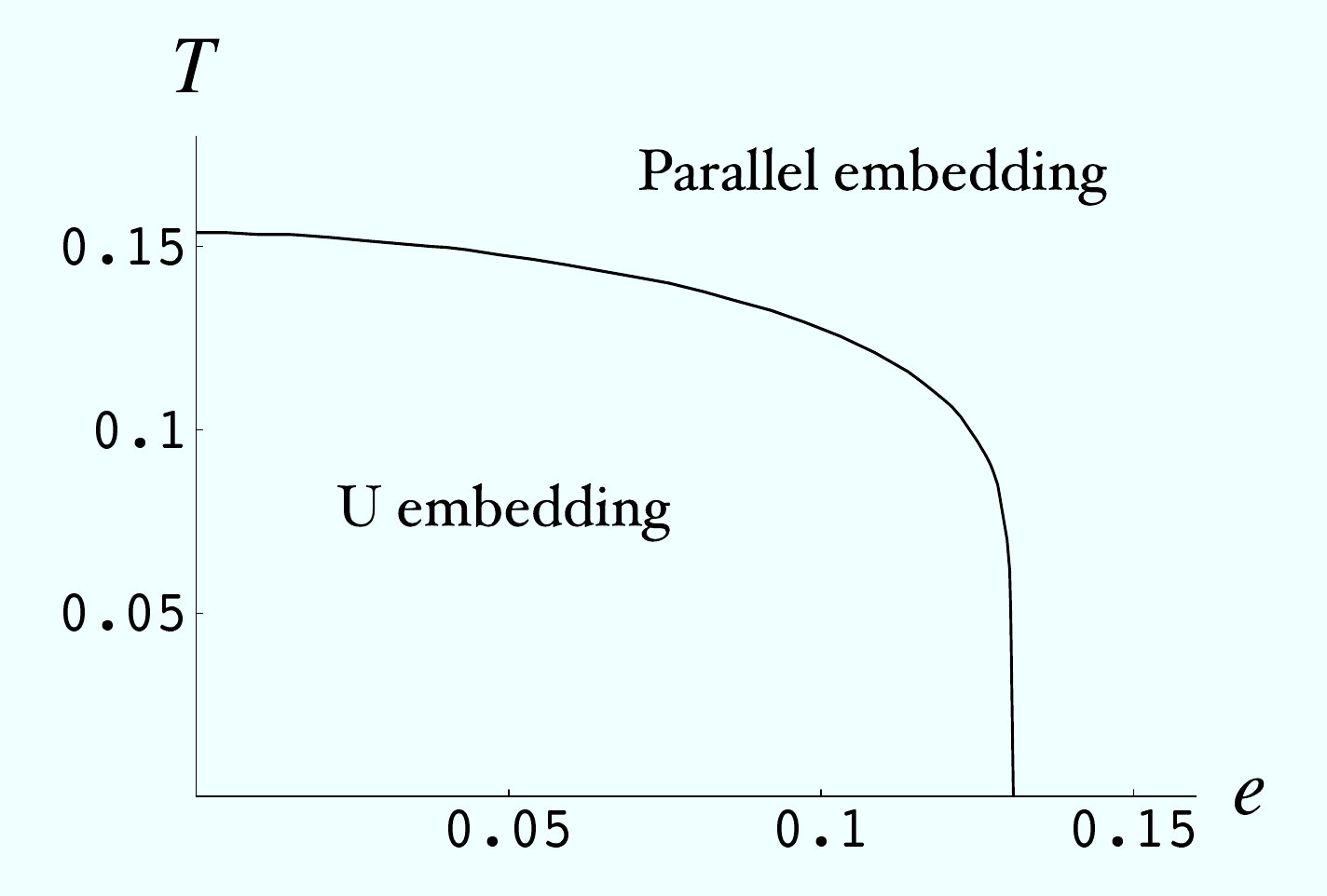, scale=0.5}
\caption{{\bf Phase diagram at nonzero temperature and electric field in the deconfining
background ($L=1$)}}
\label{electric_phase_diagram}
\end{center}
\end{figure}

Alternatively, we can get around this problem using a Maxwell-like construction
for the order parameter 
\be
c = \left.{\partial{\cal F}_e\over\partial L}\right|_{e,T} \,.
\ee
In the parallel embedding $c=0$ for any $L$.
In the U embedding the dependence of $c$ on $L$ can be determined
numerically using (\ref{L}) and (\ref{integration_constant_e}). 
Note that here we need the full solution, not just the asymptotic behavior.
The result is qualitatively the same as in the zero field case (fig.~\ref{L_c_deconfined}).
The transition occurs at the value of $L$ such that the two bounded areas are equal
(see fig.~\ref{Maxwell} for an illustration).
We can then construct the phase diagram in the $(T,e)$ plane with fixed $L$
by repeating this procedure for various values of $e$ and $T$, and finding the points
which have the same critical $L$.
The result, shown in fig.~\ref{electric_phase_diagram}, shows a first-order
insulator-conductor transition at nonzero temperature and background electric field.
At zero electric field this reduces to the chiral-symmetry breaking-restoration 
transition.

\subsubsection{Conductivity at finite density}

The conductivity computed in (\ref{vacuum_conductivity}) is for the vacuum
at a finite temperature. It has two contributions corresponding to quantum and 
thermal pair-creation of quarks and anti-quarks.
The introduction of a finite charge density should also contribute to the 
conductivity. We can compute this by generalizing our analysis in the parallel
embedding to the finite density case.
At nonzero density there is a non-trivial
time-component of the gauge field $a_0(u)$, and the associated constant
of the motion is the dimensionless baryon number charge density $d$.
The dimensionful charge density is $D=(2\pi\alpha' {\cal N}/V_4) d$.
The 8-brane action is\footnote{There is also CS term of the form
$\int a_0 a_1' (\partial_2 a_3 - \partial_3 a_2)$.
However this can be consistently set zero by the equations of motion 
since $a_0$ and $a_1$ are assumed to be independent of $x_2$ and $x_3$.}
\be
S = {\cal{N}} \int du \, u^4 \sqrt{\left(f(u) (x_4')^2 + \frac{1}{u^3} \right) 
\left(1- \frac{e^2}{f(u) u^3} \right)+ \frac{f(u)(a_1')^2}{u^3}
- \frac{(a_0')^2}{u^3}} \,.
\ee
In terms of the current and density,
the action of the parallel embedding becomes
\be
S^{||} = {\cal{N}} \int_{u_{T}}^\infty  du  \, u^{5/2} 
\sqrt{\frac{f(u) - {e^2\over u^3}}{f(u) - {j^2 - f(u) d^2\over u^5}}} \,.
\ee
The condition for reality now gives
\be
j = e\left[\left(e^2+u_T^3\right)^{2/3} + {d^2\over e^2+u_T^3}\right]^{1/2} \,,
\ee
and the conductivity is therefore
\be
\label{finite_density_conductivity}
\sigma &=& {(2\pi\alpha')^2{\cal N}\over V_4}
\left[\left(e^2+u_T^3\right)^{2/3} + {d^2\over e^2+u_T^3}\right]^{1/2} \nonumber \\
&=& {N_f N_c\lambda T^2\over 27\pi} \left[\left(1 + \tilde{e}^2\right)^{2/3} 
+ {\tilde{d}^2\over 1 + \tilde{e}^2}\right]^{1/2} \,,
\ee
where we have defined a new dimensionless (with appropriate factors of $R$,
which in our units is 1) variable $\tilde{d}$,
\be
\tilde{d} \equiv {d\over u_T^{5/2}} = {729\over 8\pi N_f N_c} {D\over\lambda^2 T^5}\,.
\ee
At zero charge density this reduces to the vacuum result (\ref{vacuum_conductivity}).


\subsection{Conductivity and the Kubo formula}

The electrical conductivity is an example of a transport coefficient that
describes the response of a thermodynamic system to a disturbance 
which takes it out of equilibrium. For a small disturbance, {\em i.e.}
near equilibrium, transport coefficients can be related to real-time correlation
functions at equlibrium via Kubo formulas.
The electrical conductivity near equilibrium is related to the current-current 
correlator \cite{CaronHuot:2006te},
\be
\sigma = \lim_{k_0\rightarrow 0}{1\over 4T} \mbox{Tr}\, C^<_{\mu\nu}(k_0=|{\bf k}|)\,,
\ee
where 
\be
C^<_{\mu\nu}(k) = \int d^4x\, e^{-ik\cdot x} \langle J_\mu(0) J_\nu(x)\rangle \,.
\ee
The correlator can in turn be computed at strong coupling using the Lorentzian 
AdS/CFT prescription of \cite{Son:2002sd}.\footnote{This prescription 
actually yields the retarded correlator
$C^R_{\mu\nu}(k) = 
i \int d^4x\, e^{-ik\cdot x} \theta(x_0)\langle [J_\mu(0),J_\nu(x)]\rangle$,
which is related to the one above by
$C^<_{\mu\nu}(k) = 2\mbox{Im}\, C^R_{\mu\nu}(k)/(e^{-k_0/T} - 1)$.}
It is therefore interesting to compare the result of this computation
with the direct computation of the conductivity in the previous section.
Since the Kubo formula gives the conductivity near equilibrium,
one should compare the result with the zero electric field limit of
(\ref{finite_density_conductivity}).

The current-current correlator for light-like momenta has been analyzed
in the Sakai-Sugimoto model at both finite temperature and density 
in \cite{Parnachev:2006ev}. 
In particular the high-temperature, or equivalently low frequency, behavior
was found to be\footnote{Our definition of $\lambda$ is different than in
\cite{Parnachev:2006ev}: $\lambda_{here} = 4\pi\lambda_{there}$.
Our $\tilde{D}$ is exactly their $\tilde{C}$.}
\be
\lim_{k_0\rightarrow 0} \mbox{Tr}\, C^<(k_0=|{\bf k}|) = 
{2 (2N_f) N_c \lambda T^3\over 27\pi}
\sqrt{1 + \tilde{D}^2}\,,
\ee
leading to a conductivity
\be
\sigma = {N_f N_c \lambda T^2\over 27\pi}
\sqrt{1 + \tilde{D}^2}\,,
\ee
in perfect agreement with the zero field limit of (\ref{finite_density_conductivity}).


\subsection{Confined phase}
\begin{figure}[htbp]
\begin{center}
\begin{tabular}{cc}
\epsfig{file=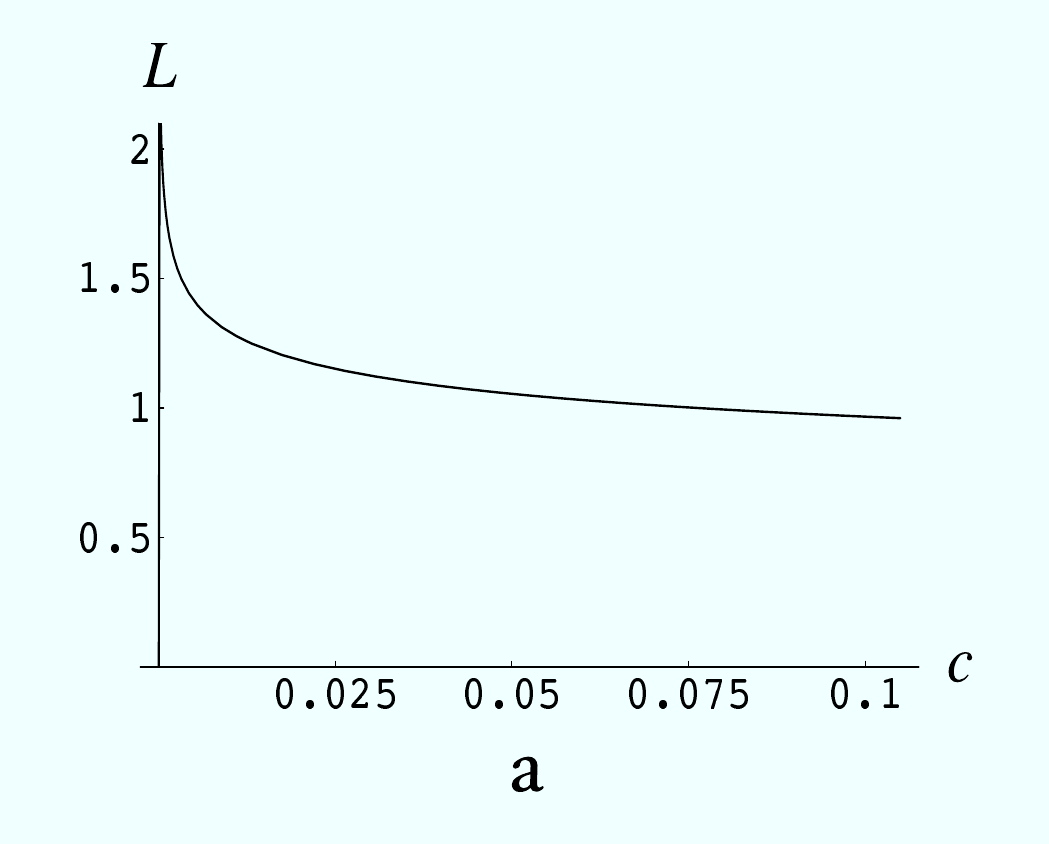,scale=0.55} &
\epsfig{file=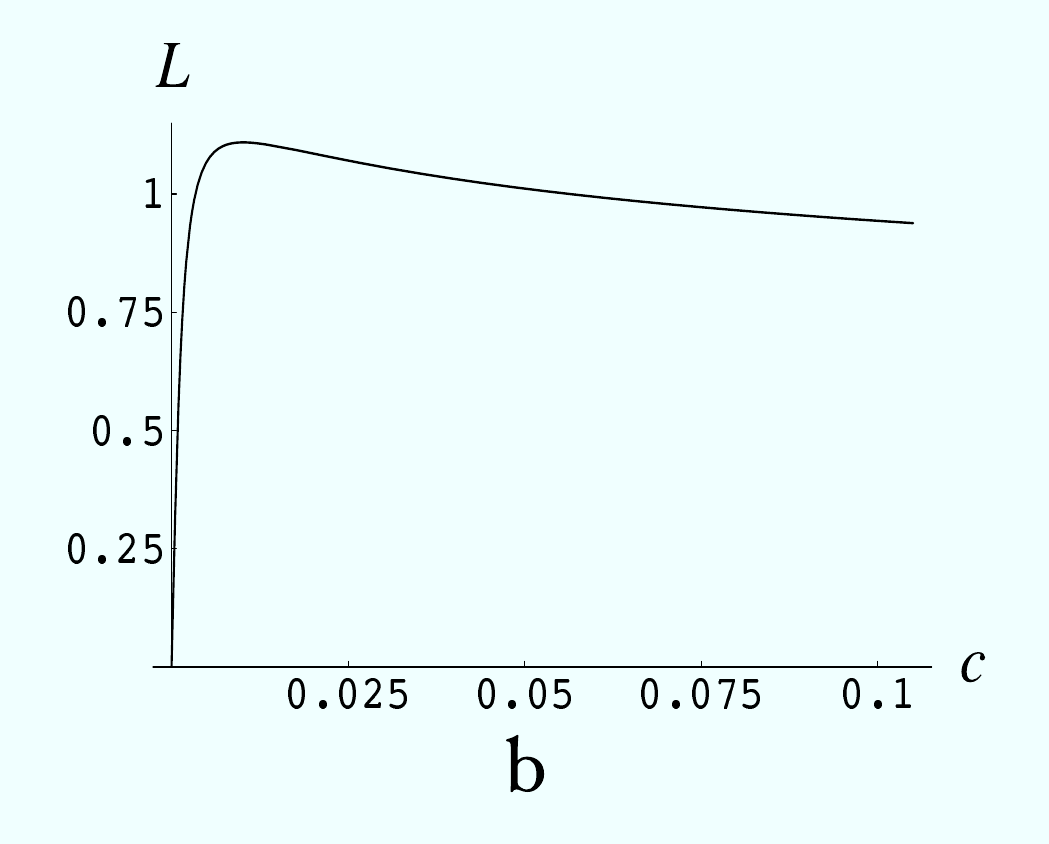,scale=0.55}
\end{tabular}
\caption{{\bf L vs. c in the confined phase (a) $e < u_{KK}^{3/2}$
(b) $e>u_{KK}^{3/2}$}}
\label{L_vs_c_confined}
\end{center}
\end{figure}

For $T<1/(2\pi R_4)$ the confining background dominates, and the 8-brane
action with the background gauge field is
\be
S &=& {\cal{N}} \int du \, u^4 \sqrt{\left(f(u) (x_4')^2 + \frac{1}{f(u) u^3} \right) 
\left(1- \frac{e^2}{u^3} \right) + \frac{(a_1')^2}{u^3}} \nonumber\\
&=& {\cal{N}} \int du \, u^4 \sqrt{\left(f(u) {x_4'}^2 + \frac{1}{f(u) u^3} \right) 
\left(1- \frac{e^2}{u^3} \right)\left(1-{j^2\over u^5}\right)^{-1}} \,.
\ee
At zero field the only possible embedding in the confined
phase was the U embedding.
However as we increase the electric field we encounter a puzzle.
As in the deconfined phase, the U embedding solution at a fixed $L$ 
ceases to exist above a certain value of the field.
To see this we again study the behavior of $L$ as a function of $c$.
The U embedding with $j=0$ satisfies
\be
\label{U_confined_e}
x_4'(u) = {1\over u^{3/2} f(u)} 
\left[{u^8 f(u)\left(1-{e^2\over u^3}\right)
\over u_0^8 f(u_0) \left(1 - {e^2\over u_0^3}\right)} - 1\right]^{-1/2} \,,
\ee
and
\be
\label{integration_constant_confined_e}
c = u_0^4 \sqrt{f(u_0)\left(1-{e^2\over u_0^3}\right)} \,.
\ee
There are two cases to consider.
For $e^2<u_{KK}^3$, $L(c)$ decreases monotonically 
from $L(0)=\pi R_4$ (the anti-podal embedding) to zero as $c\rightarrow\infty$
(fig.~\ref{L_vs_c_confined}a).
For $e^2>u_{KK}^3$
the asymptotic behavior of $L(c)$ is the same as in the deconfined phase
(fig.~\ref{L_vs_c_confined}b),
implying a maximal $L$ for a given $e$,
or alternatively a maximal $e$ for a given $L$.
This maximal value therefore satisfies $u_{KK}^3<e^2_{max}<u_0^3$.
The U embedding exists only for $e<e_{max}$, and its action is given by
\be
\label{electric_confined_U_action}
S^U = {\cal{N}} \int_{u_0}^\infty  du  \, {u^{5/2}\over \sqrt{f(u)}}
\sqrt{1-{e^2\over u^3}}
\left[1 - {u_0^8 f(u_0) \left(1-{e^2\over u_0^3}\right)
\over u^8f(u)  \left(1 - {e^2\over u^3}\right)}\right]^{-1/2} \,.
\ee

\begin{figure}[htbp]
\begin{center}
\epsfig{file=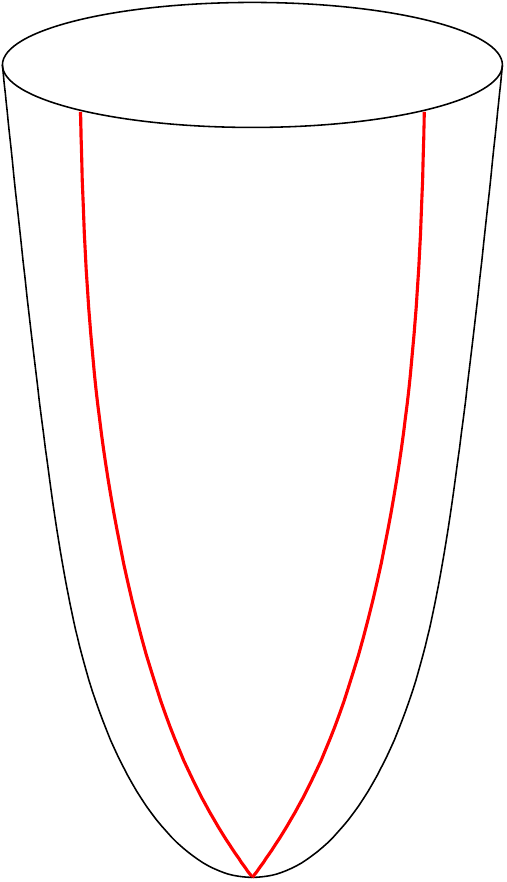, scale=0.4}
\caption{{\bf The conducting V embedding in the confining background}}
\label{V_embedding}
\end{center}
\end{figure}

The question is what happens when $e>e_{max}$?
There must exist a second embedding that takes over before reaching the maximal
electric field. In the deconfined background this was the current-carrying 
parallel embedding, and we observed a first-order phase transition at 
$e<e_{max}$. We therefore propose a new kind of 8-brane embedding
in the confining background, which is analogous to the parallel embedding in the
deconfining background. The 8-brane and anti-8-brane follow parallel radial 
geodesics and connect at $u=u_{KK}$ (fig.~\ref{V_embedding}).
In this ``V-shaped" embedding $x_4'(u)=0$ (and therefore $c=0$) except at the tip, 
where there is a cusp
(unless we are in the anti-podal embedding, which is smooth).
Away from $u_{KK}$ this is clearly a solution.
Its action is given by
\be
S^V = {\cal N}\int_{u_{KK}}^\infty du\, {u^{5/2}\over \sqrt{f(u)}}
\sqrt{{1-{e^2\over u^3}}\over 1-{j^2\over u^5}}\,.
\ee
It follows from reality of the action that if $e^2>u_{KK}^3$ there must be a 
current given by
$j = e^{5/3}$. This embedding is therefore a conductor, with a conductivity
\be
\label{confined_conductivity}
\sigma = {(2\pi\alpha')^2{\cal N}\over V_4}\, e^{2/3} =
{N_f N_c\over 12 \pi^{7/3}}\, \lambda^{1/3} E^{2/3} \,.
\ee

This proposal raises two questions.
The first has to do with stability of the cusp singularity at the tip, and the second
with the identification of the current carriers.
In the deconfined parallel embedding the current is carried by 
fundamental strings, namely quarks and anti-quarks.
The only charged objects in the confined phase
are baryons, so the current can only be carried by baryons and anti-baryons,
namely by 4-branes (and anti-4-branes) wrapped on the $S^4$.
Indeed, this is precisely what we need to source the 5d magnetic field
$a_1'(u)$ dual to the boundary current $j$.
This is similar to the situation at nonzero density \cite{bll},
where a uniform distribution of 4-branes was used
to source the 5d electric field $a_0'(u)$ dual to the baryon number density.
In that case the 4-branes created a cusp in the 8-brane embedding.
In the present case the current corresponds to a uniform distribution
of 4-branes and anti-4-branes (since the total charge density vanishes),
located at the cusp, and moving at a constant velocity along $x_1$.
To understand whether this configuration is stable we need to understand
the forces at the cusp. In the nonzero density case the upward force due to the 8-brane
was balanced against the downward force of the 4-branes.
In this case however the 4-branes are at the bottom of the space, and it is not clear
how the balance comes about.
Higher derivative corrections may also be relevant.
We leave this as an open question.

\begin{figure}[htbp]
\begin{center}
\begin{tabular}{cc}
\epsfig{file=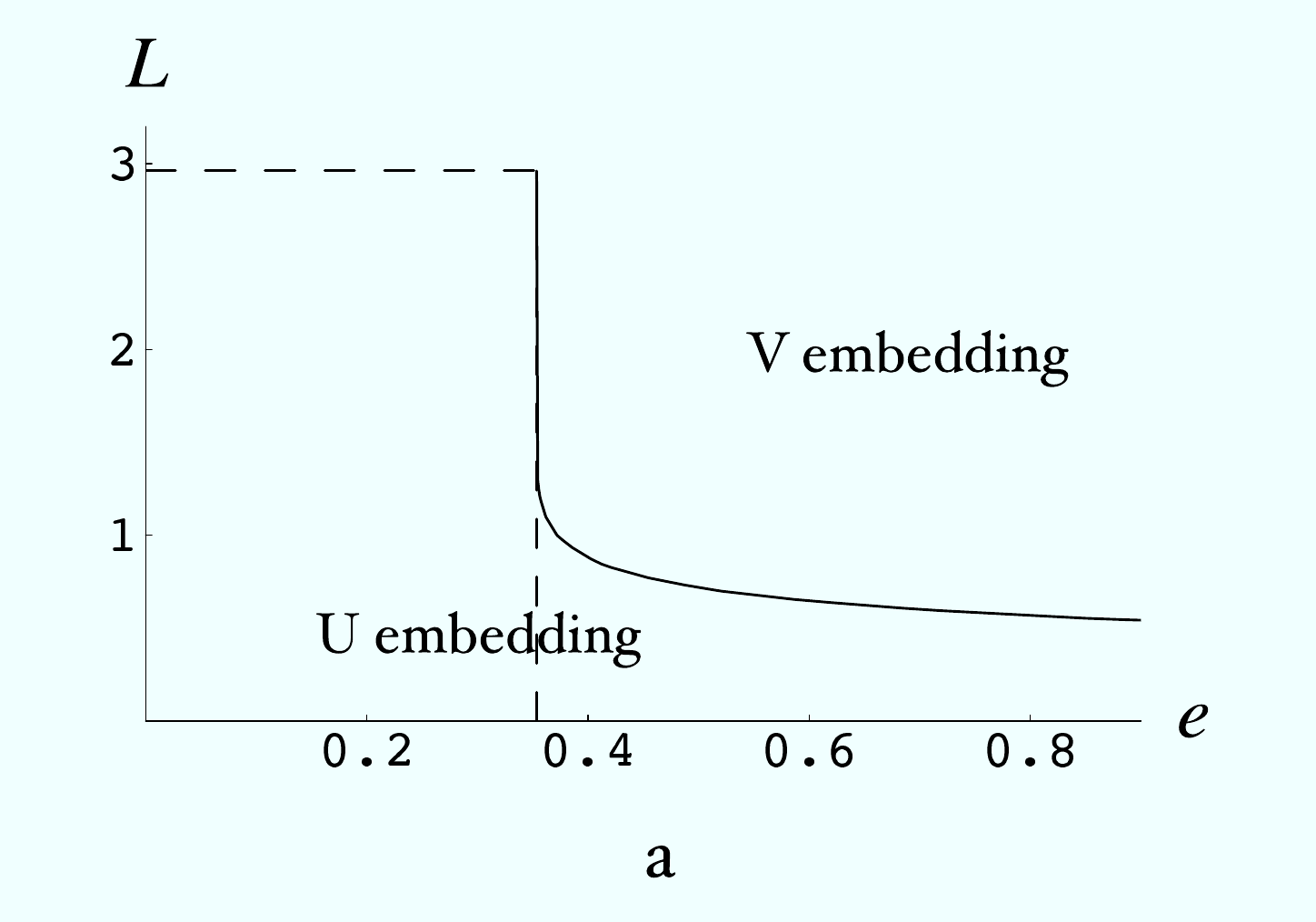,scale=0.5} &
\epsfig{file=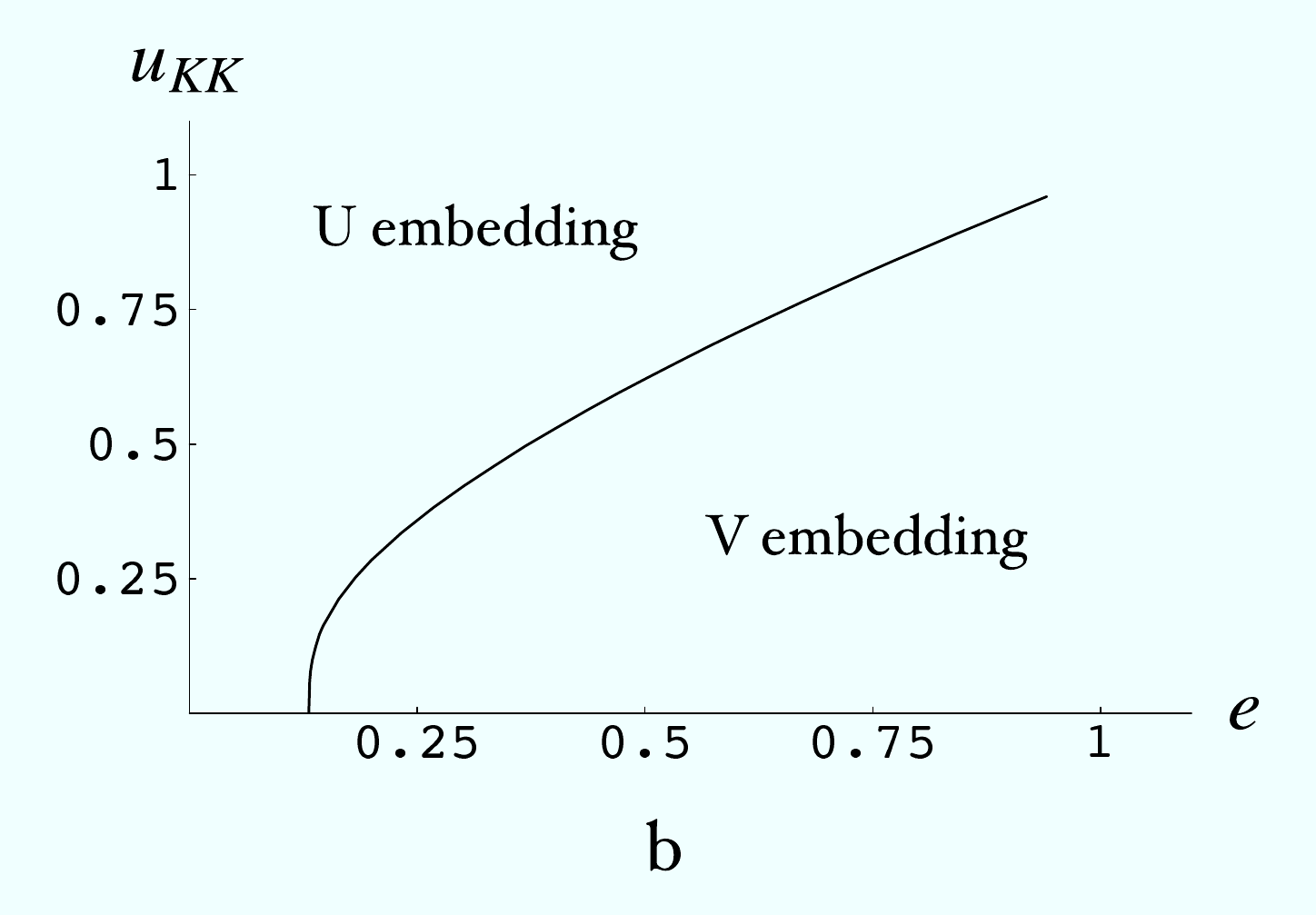,scale=0.5}
\end{tabular}
\caption{{\bf Electric phase diagram in the confined phase 
(a) fixed $u_{KK}=0.5$ (b) fixed $L=1$.  In (a) the dashed lines illustrate that 
for $e=u_{KK}^{3/2}\sim 0.35$ the critical $L=\pi R_4 \sim 3$.}}
\label{electric_confined_phase_diagram}
\end{center}
\end{figure}

Assuming this is a valid solution, we can construct the phase diagram
using the same method as in the deconfined phase.
In this case we can either fix $u_{KK}$ and vary $L$ and $e$,
or fix $L$ and vary $u_{KK}$ and $e$. The two results are shown in
figure \ref{electric_confined_phase_diagram}.
From the first diagram we see that 
for $e^2<u_{KK}^3$ the U embedding dominates at all values of $L$,
and that for $e^2\geq u_{KK}^3$ there is a first-order transition to
the conducting V embedding at a critical $L$ that starts at $\pi R_4$
(the anti-podal embedding) and decreases with $e$.

The second diagram also provides some interesting insight.
First, we note that, for a fixed $L$, the critical electric field in the limit
$u_{KK}\rightarrow 0$ in fig.~\ref{electric_confined_phase_diagram}b
is exactly the same as the critical field in the deconfined phase 
in the limit $T\rightarrow 0$ (as seen in fig.~\ref{electric_phase_diagram}).
This is because both cases are then essentially equivalent to the non-compact
model at zero temperature. 
Second, by comparing the phase diagrams in the confined and deconfined phases 
at the confinement/deconfinement transition $u_{KK}=u_T$,
it can be seen that if we start in the confined 
conducting phase (V embedding) and raise the temperature above the deconfinement
temperature we end up in the deconfined conducting phase (parallel embedding).
In this transition the conductivity jumps from (\ref{confined_conductivity})
to (\ref{vacuum_conductivity}).


\subsection{Electric susceptibility}

\begin{figure}[htbp]
\begin{center}
\begin{tabular}{cc}
\epsfig{file=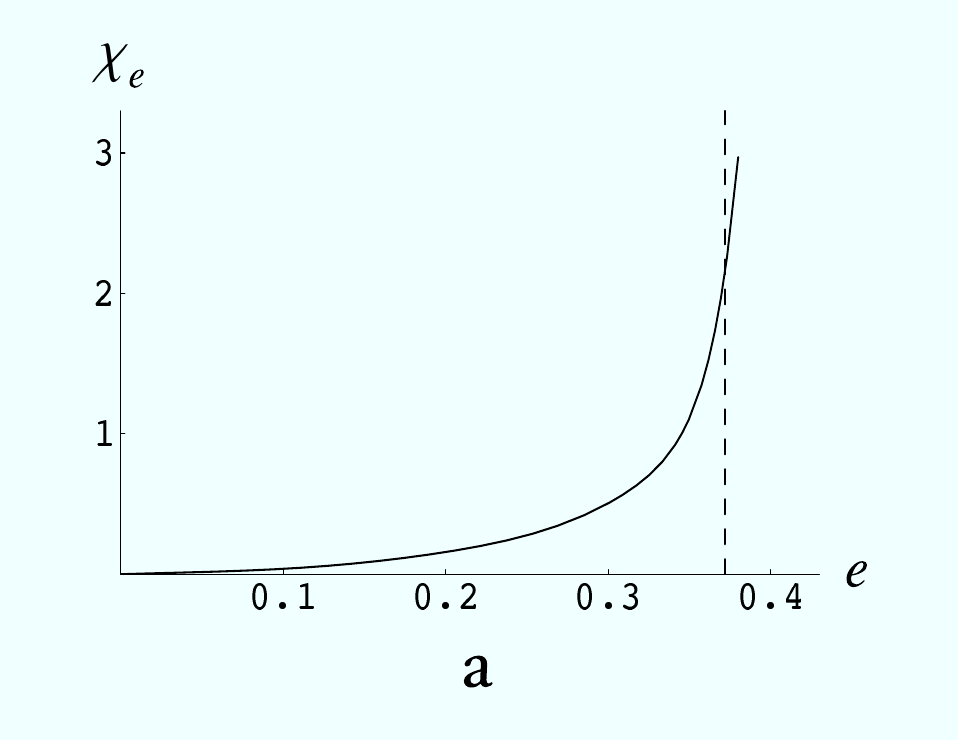,scale=0.65} &
\epsfig{file=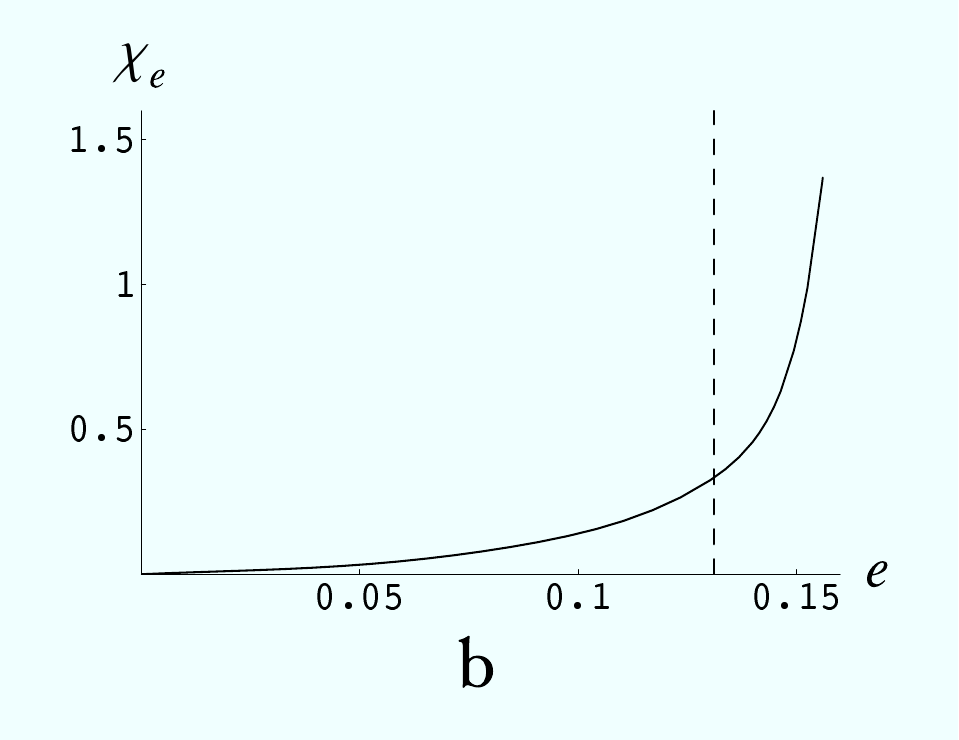,scale=0.65}
\end{tabular}
\caption{{\bf Electric susceptibility in the insulating phase of the (a) confined phase,
(b) deconfined phase (the dashed line shows the critical field for the
insulator/conductor transition).}}
\label{electric_susceptibility_figure}
\end{center}
\end{figure}

In both the confined and deconfined phases there is an insulating phase
(U embedding) in which the current vanishes.
These phases exhibit an electric polarization and susceptibility,
which are defined thermodynamically by
\be
p = -{\partial{\cal F}_e\over \partial e} \; , \;
\chi_e = -{\partial^2{\cal F}_e\over \partial e^2} \,.
\ee
In the holographic prescription the free energy is divergent, and requires
a counterterm proportional to $e^2$.
The susceptibility therefore requires a counterterm independent of $e$.
A physically motivated scheme is to require the susceptibility to vanish at zero field,
in other words
\be 
\label{electric_susceptibility}
\chi_e(e) = - {\partial^2{\cal F}_e\over \partial e^2} +
\left.{\partial^2{\cal F}_e\over \partial e^2}\right|_{e=0} \,.
\ee
This is a measure of the nonlinearity of the vacuum in this model.
The results are presented in figure \ref{electric_susceptibility_figure}.


\section{Magnetic field}

For an external magnetic field $H$ our ansatz is simply
\be
\hat{A}_2(x_1) = Hx_1 \,.
\ee
There will be no current and therefore no $u$ dependence.
Our main objective here is to determine the effect of the background magnetic
field on the critical temperature for chiral-symmetry restoration in the
deconfined phase. 
We will also compute the magnetic susceptibility of the vacuum in both the 
confined and deconfined phases.


\subsection{Deconfined phase}

\begin{figure}[htbp]
\begin{center}
\begin{tabular}{cc}
\epsfig{file=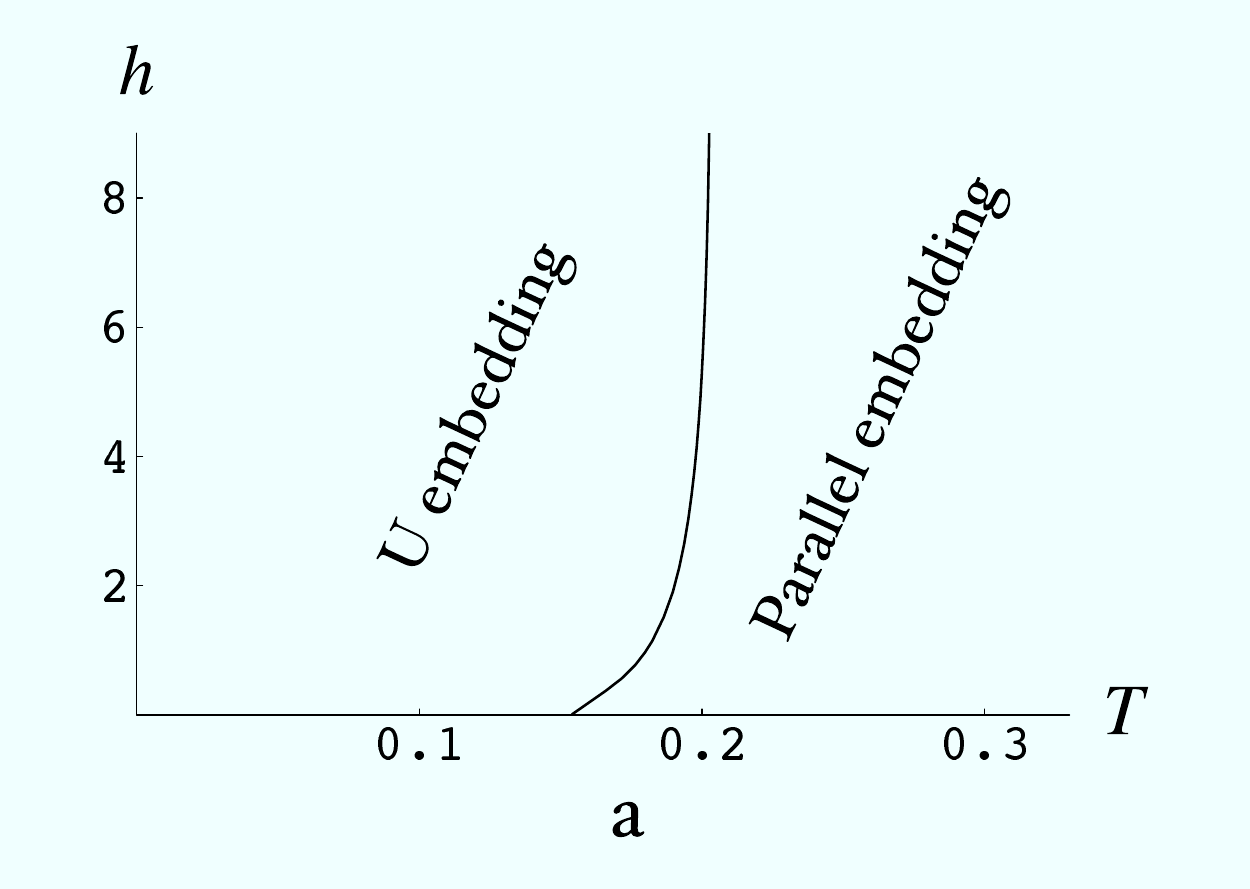,scale=0.5} &
\epsfig{file=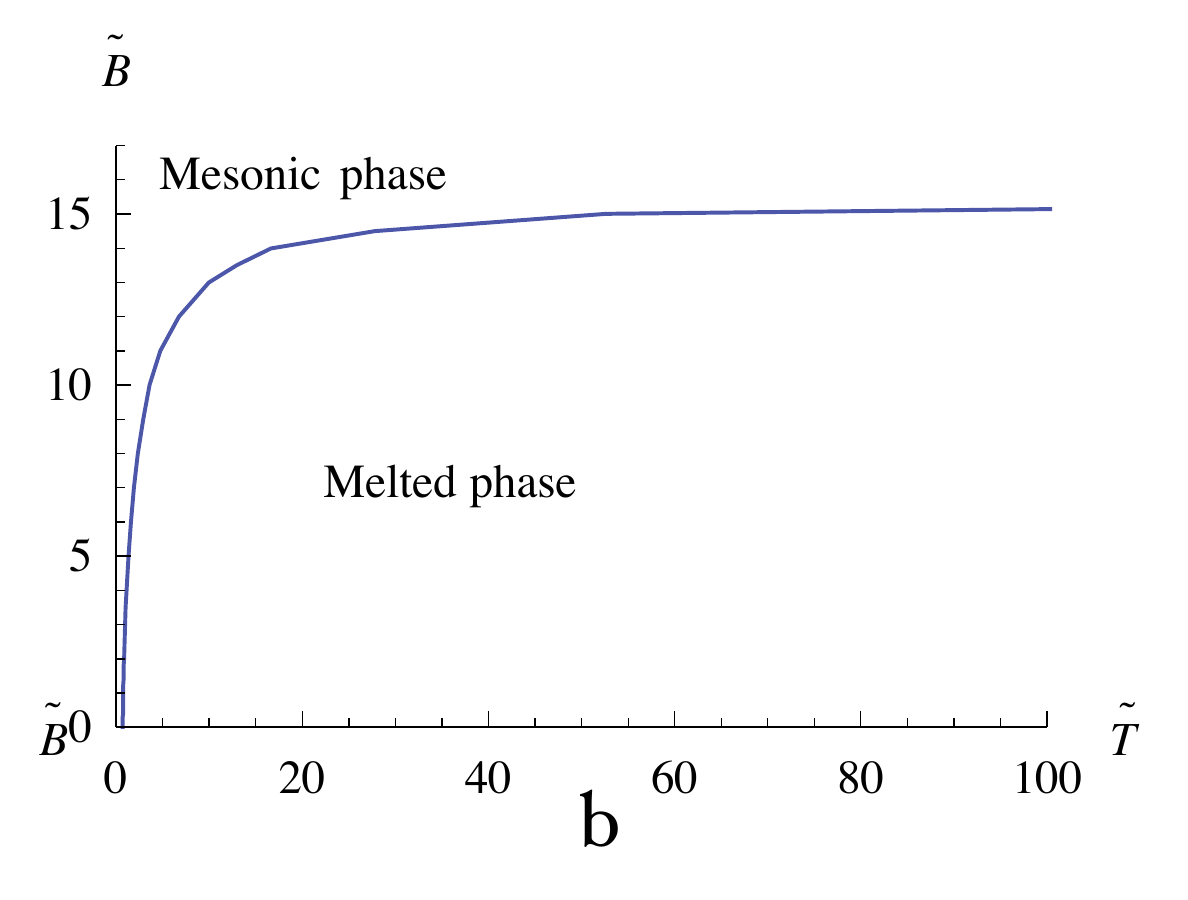,scale=0.5}
\end{tabular}
\caption{{\bf (a) Our phase diagram with a magnetic field
(b) analogous phase diagram in the D3-D7 model 
(reprinted from \cite{Erdmenger} with the authors' permission).}}
\label{magnetic_phase_diagram}
\end{center}
\end{figure}

We start again in the deconfined phase.
The 8-brane action in the magnetic field background is given by
\be 
\label{D8_action_deconfined}
S = {\cal{N}} \int du \, u^4 \sqrt{\left(f(u) {x_4'}^2 + \frac{1}{u^3}\right) \left(1+ \frac{h^2}{u^3} \right)} \,,
\ee
where $h\equiv 2\pi\alpha' H$.
The chiral symmetry breaking U embedding now satisfies
\be
\label{U_deconfined_m}
x_4'(u) = {1\over u^{3/2} \sqrt{f(u)}} 
\left[{u^8\left(f(u)+{h^2\over u^3}\right)
\over u_0^8 \left(f(u_0) + {h^2\over u_0^3}\right)} - 1\right]^{-1/2} \,,
\ee
and has an action
\be
\label{deconfined_U_solution_m}
S^{U} = {\cal{N}} \int_{u_0}^\infty  du  \, u^{5/2}
\sqrt{1+{h^2\over u^3 f(u)}}
\left[1 - {u_0^8\left(f(u_0)+{h^2\over u_0^3}\right)
\over u^8 \left(f(u) + {h^2\over u^3}\right)}\right]^{-1/2} \,.
\ee
In the chiral-symmetric parallel embedding $x_4'(u)=0$, and the action is
\be
S^{||} = {\cal{N}} \int_{u_{T}}^\infty  du  \, u \sqrt{u^3 + h^2} \,.
\ee
The actions of the embeddings, which define the magnetic free energies
${\cal F}_m(L,T,h)$, are divergent, but the difference is finite.
The resulting phase diagram (for a fixed value of $L$) is shown in
figure \ref{magnetic_phase_diagram}a.
Note that unlike the electric field case, here both embeddings describe equilibrium
states, so we can compare the actions directly.

We observe that the temperature at which chiral symmetry is restored
increases with the background magnetic field, and approaches a finite value
in the limit of an infinite field.
This means that above some nonzero temperature chiral symmetry is always restored.
A similar increase in the critical temperature for the phase transition in the D3-D7
model was observed in \cite{Johnson,Erdmenger}, 
but there is a crucial difference with our result.
In the D3-D7 model the critical temperature
diverges at a finite value of the magnetic field, which means
that there is no phase transition for magnetic fields larger than this value
(fig.~\ref{magnetic_phase_diagram}b).
It is ammusing to speculate whether this can be tested in real QCD,
either experimentally or on the lattice.


\subsection{Confined phase}

In the confined phase the 8-brane action is
\be 
S =  {\cal{N}} \int du \, u^4 \sqrt{\left(f(u) {x_4'}^2 + \frac{1}{f(u) u^3}\right) 
\left(1+ \frac{h^2}{u^3}\right)} \,.
\ee
The U embedding is basically the same as in the electric case,
(\ref{U_confined_e}) and (\ref{integration_constant_confined_e}),
with $e^2$ replaced by $-h^2$.
However, the sign difference guarantees that this solution is the only one
and that it exists for all values of $L$ and $h$.
The action of the solution can also be read off from the electric case
(\ref{electric_confined_U_action}) with the above replacement,
\be
\label{magnetic_confined_U_action}
S^U = {\cal{N}} \int_{u_0}^\infty  du  \, {u^{5/2}\over \sqrt{f(u)}}
\sqrt{1+{h^2\over u^3}}
\left[1 - {u_0^8 f(u_0) \left(1+{h^2\over u_0^3}\right)
\over u^8f(u)  \left(1 + {h^2\over u^3}\right)}\right]^{-1/2} \,.
\ee

\subsection{Magnetic susceptibility}
\begin{figure}[htbp]
\begin{center}
\begin{tabular}{ccc}
\epsfig{file=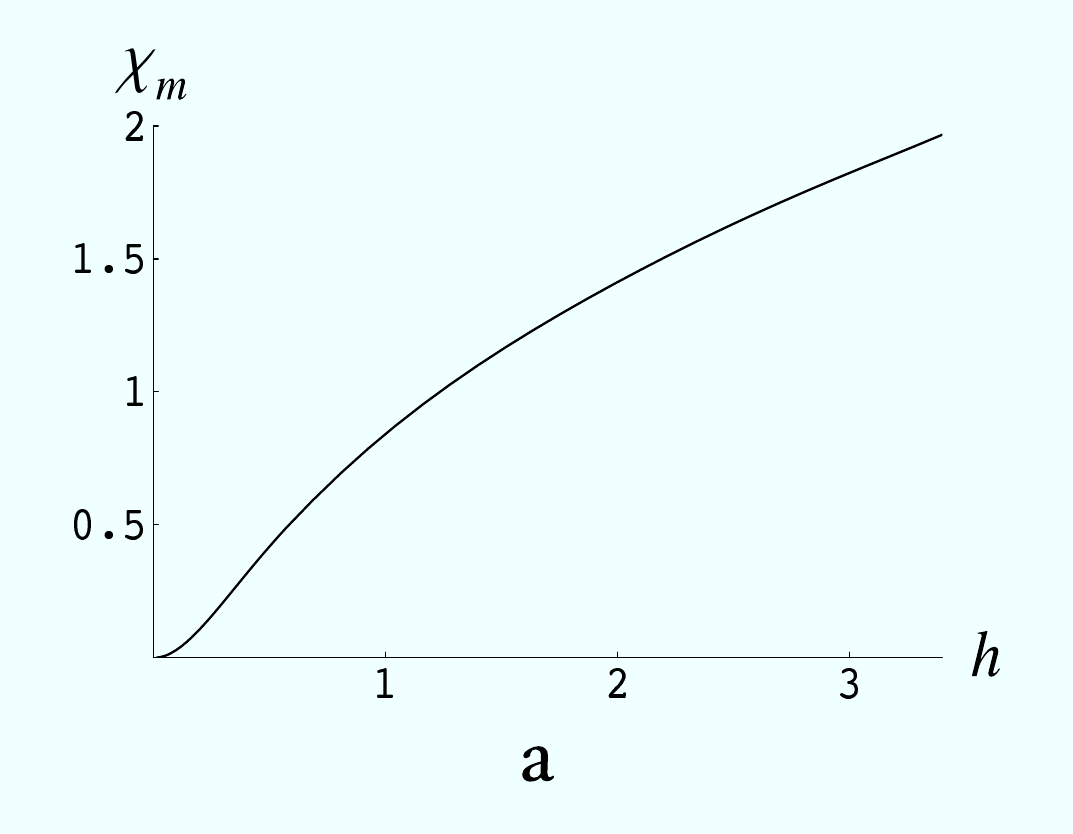,scale=0.45} &
\epsfig{file=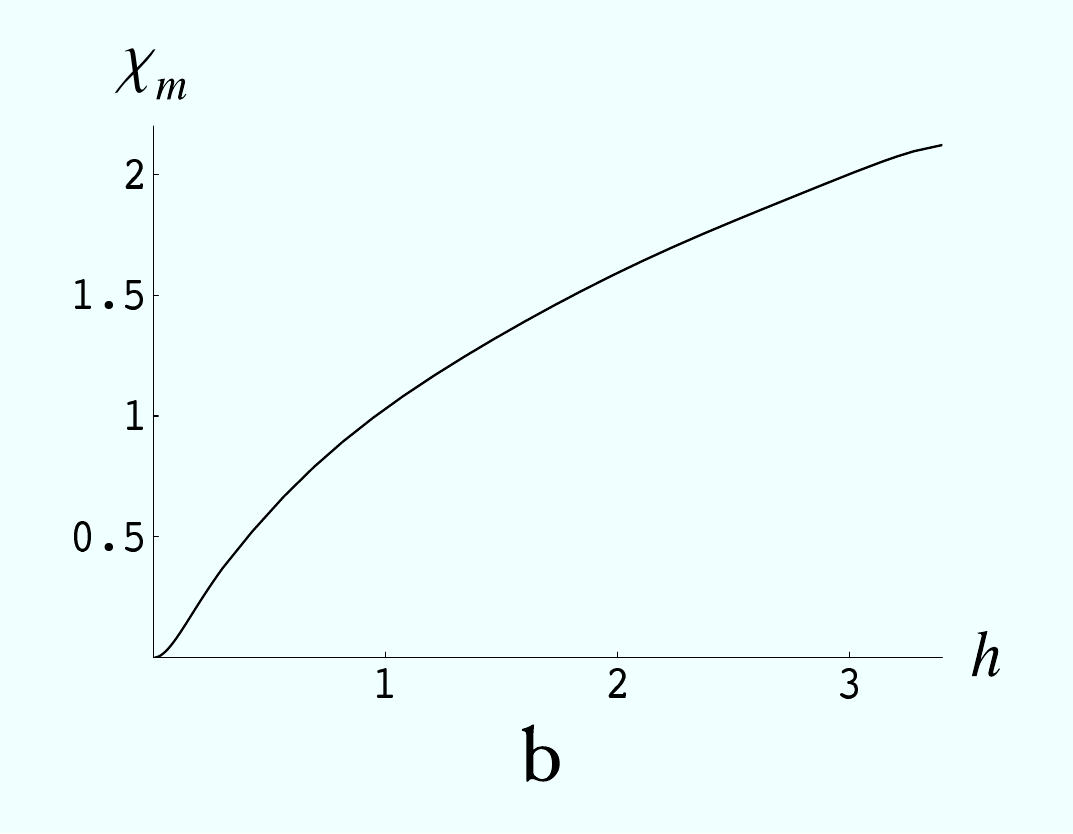,scale=0.45} &
\epsfig{file=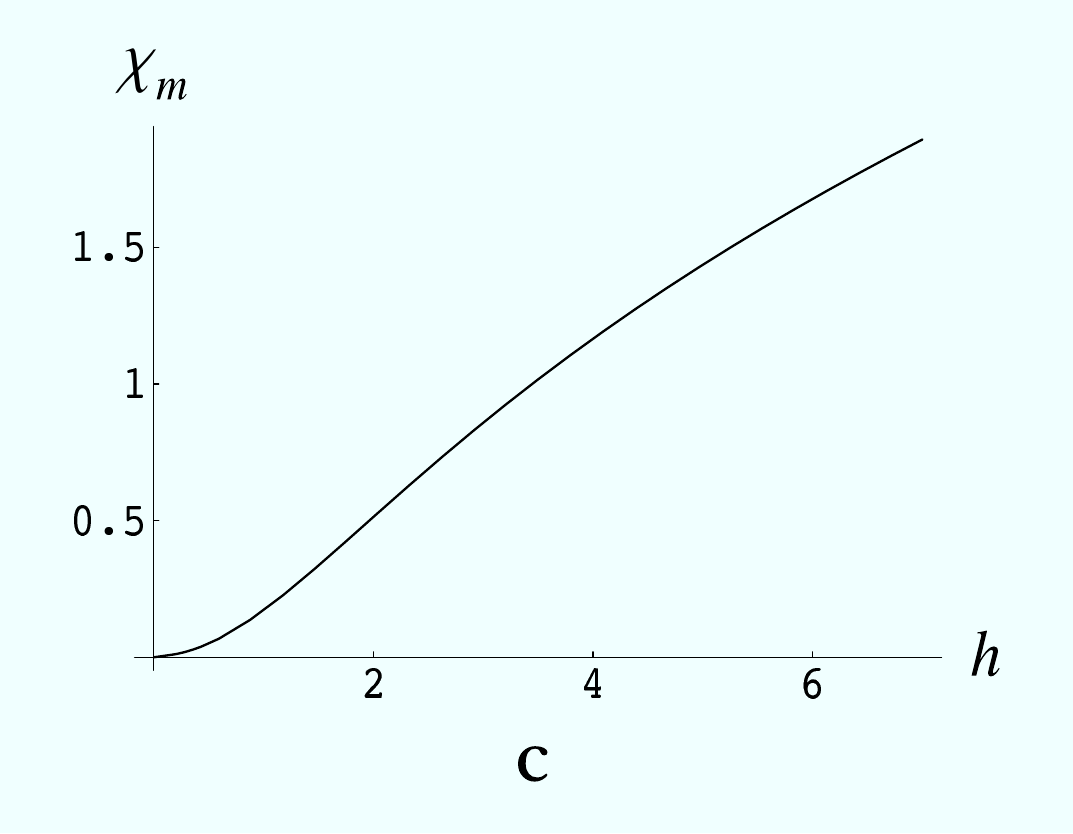,scale=0.45}
\end{tabular}
\caption{{\bf Magnetic susceptibility: (a) confined, broken chiral symmetry
(b) deconfined, broken chiral symmetry (c) deconfined, restored chiral symmetry}}
\label{magnetic_susceptibility_figure}
\end{center}
\end{figure}

The magnetization and magnetic susceptibility are defined thermodynamically by
\be
m = -{\partial{\cal F}_m\over \partial h} \; , \;
\chi_m = -{\partial^2{\cal F}_m\over \partial h^2} \,.
\ee
We can compute the deviation from linearity by regularizing
the susceptibility as in the electric case,
\be 
\label{magnetic_susceptibility}
\chi_m(h) = - {\partial^2{\cal F}_m\over \partial h^2} +
\left.{\partial^2{\cal F}_m\over \partial h^2}\right|_{h=0} \,.
\ee
The results for all three phases are shown in figure \ref{magnetic_susceptibility_figure}.


\section*{Acknowledgments}
We would like to thank Ren\'e Meyer and Johanna Erdmenger for 
useful correspondence. 
This work was supported in part by the
Israel Science Foundation under grant no.~568/05.
OB also gratefully acknowledges support from the Institute for Advanced Study.

\end{document}